\documentclass{article}
\usepackage{cite}
\usepackage{amsmath}
\begin{document}
\begin{titlepage}
\vspace*{10mm}
\begin{center}
{\large\bfseries A family of discrete-time exactly-solvable
exclusion processes on a one-dimensional lattice}\\
\vspace*{\baselineskip}

{\textbf{Farinaz~Roshani} $^{1,a}$ and \textbf{Mohammad~Khorrami}
$^{2,b}$}\\
\vspace*{\baselineskip}

{$^1$ \textit{Institute for Advanced Studies in Basic Sciences,}}
\\ {\textit{P. O. Box 159, Zanjan 45195, Iran}}\\
{$^2$ \textit{Department of Physics, Alzahra University,}}
\\ {\textit{Tehran, 19938-91167, Iran}}\\
$^a$ farinaz@iasbs.ac.ir\\
$^b$ mamwad@mailaps.org\\
\end{center}
\vspace*{3\baselineskip}
\begin{abstract}
\noindent A two-parameter family of discrete-time exactly-solvable
exclusion processes on a one-dimensional lattice is introduced,
which contains the asymmetric simple exclusion process and the
drop-push model as particular cases. The process is rewritten in
terms of boundary conditions, and the conditional probabilities
are calculated using the Bethe-ansatz. This is the discrete-time
version of the continuous-time processes already investigated in
\cite{AKK1,AKK2,RK}. The drift- and diffusion-rates of the
particles are also calculated for the two-particle sector.
\end{abstract}
\vspace*{\baselineskip} PACS numbers: 05.40.-a, 02.50.Ga
\end{titlepage}

\section{Introduction}
The asymmetric exclusion process and the problems related to it,
including for example bipolymerization \cite{4}, dynamical models
of interface growth \cite{5}, traffic models \cite{6}, the noisy
Burgers equation \cite{7}, and the study of shocks \cite{8,9},
have been extensively studied in recent years. The dynamical
properties of this model have been studied in \cite{9,11}. As the
results obtained by approaches like mean field are not reliable in
one dimension, it is useful to introduce solvable models and
analytic methods to extract exact physical results.

The totally asymmetric simple exclusion model on a one-dimensional
lattice is one of the simplest examples from which exact results
can be obtained. Such systems consist of a lattice in which every
site is either empty or occupied by a single particle. Particles
can hop to the right, if their right-neighboring site is empty.
The steady-state of such systems have been extensively studied,
for continuous-time as well as discrete-time evolutions. Among the
methods used to study the steady-state properties of such systems
is the matrix-product ansatz, \cite{A1,A2,A3,A4,A5}. Various
methods have also been used to study the time-dependent state of
such systems. In \cite{A6,A7,A8}, generalizations of the
matrix-product ansatz have been used to study asymmetric exclusion
processes. In \cite{GS}, an explicit form for the conditional
probability of finding particles on specific sites in a system of
asymmetric exclusion process was obtained in terms of a
determinant.

In \cite{GS}, the coordinate Bethe-ansatz is used to solve the
asymmetric simple exclusion process on a one-dimensional lattice.
In \cite{AKK1}, a similar technique was used to solve the
drop-push model, and a generalized one-parameter model
interpolating between the asymmetric simple exclusion model and
the drop-push model. In \cite{AKK2}, this family was further
generalized to a family of processes with arbitrary left- and
right- diffusion rates. All of these models were lattice models.
The behaviour of the latter model on a continuum was investigated
in \cite{RK}. The discrete-time version of the asymmetric
exclusion process was discussed in \cite{BPS}. In
\cite{AA,RK3,RK4}, a similar Bethe-ansatz approach was used to
study exclusion systems consisting of several kinds of particles.

Here we consider discrete-time asymmetric exclusion processes in a
one-dimensional lattice. The scheme of the paper is the following.
In section 2, a system is introduced which consists of a
one-dimensional lattice in which each of the sites are either
empty or occupied by a single particle. A discrete-time evolution
is introduced and it is shown that the interaction between
particles can be substituted by a suitable boundary condition. In
section 3, the conditional probability of occupied sites is
obtained. In section 4, the drift rates for the two particle
sector are calculated. In section 5, the diffusion rate for the
two particle sector is calculated. Section 6 is devoted to the
concluding remarks.

It is seen that for large times, the results of the
continuous-time evolution are recovered, namely that the drift
rates tend to the no-interaction drift rates, while the diffusion
rate is generally larger than the diffusion rate of the
non-interacting system.

\section{A family of discrete-time exclusion processes on a one-dimensional lattice}
Consider a one-dimensional lattice, in which each site is either
empty or occupied by one particle. The probability that the first
particle is in $x_1$, the second particle is in $x_2$, etc. is
denoted by
$$P(x_1,x_2,\dots), \qquad x_1<x_2<\cdots $$
The process is that each particle can hop to the right, with the
probability $\alpha$, if the its right-hand side neighbor is
empty:
\begin{equation}\label{f1}
A\emptyset\to\emptyset A,\qquad\hbox{with the probability
}\alpha
\end{equation}
Consider the following evolution equation and boundary condition
for the two-particle sector.
\begin{align}\label{f2}
P(x_1,x_2,t+1)=&(1-\alpha)^2\,P(x_1,x_2,t)\nonumber\\
&+\alpha\,(1-\alpha)\,[P(x_1-1,x_2,t)+P(x_1,x_2-1,t)]\nonumber\\
&+\alpha^2\,P(x_1-1,x_2-1,t),\qquad x_1<x_2,
\end{align}
and
\begin{equation}\label{f3}
P(x,x)=\lambda\,P(x,x+1)+\mu\,P(x-1,x),\qquad \lambda+\mu=1.
\end{equation}
Eq. (\ref{f2}) describes a system with a diffusion process which
occurs simultaneously for all particles. This is in contrast to a
system for which at each step only one particle can hop to the
right (if its right-hand site is empty). In the latter case, terms
proportional to $\alpha^2$ would be omitted from the above
equation.

Using (\ref{f3}), it is seen that
\begin{align}\label{f4}
P(x,x+1,t+1)=&[(1-\alpha)^2+\lambda\,\alpha\,(1-\alpha)]\,P(x,x+1,t)\nonumber\\
&+\alpha\,(1-\alpha)\,P(x-1,x+1,t)\nonumber\\&+[\alpha^2+\mu\,\alpha\,(1-\alpha)]
\,P(x-1,x,t).
\end{align}
So it is seen that (\ref{f2}) and (\ref{f3}) describe a system
where particles can push:
\begin{equation}\label{f5}
AA\emptyset\to\emptyset AA,\qquad\hbox{with the probability }\beta
\end{equation}
where
\begin{equation}\label{f6}
\beta=\mu\,\alpha\,(1-\alpha)+\alpha^2=\alpha-\lambda\,\alpha\,(1-\alpha).
\end{equation}

One can use (\ref{f2}) and (\ref{f3}), to obtain pushing rates in
multi-particle sectors as well. This is especially simple in two
cases: $(\lambda=1,\mu=0)$ and $(\lambda=0,\mu=1)$. In the first
case, one obtains
\begin{align}\label{f7}
P(x,x+1,\dots,x+n,t+1)=&(1-\alpha)\sum_{m=0}^n\alpha^m\nonumber\\
&\times P(\dots,x+m-2,x+m,\dots,x+n,t)\nonumber\\
&+\alpha^{n+1}\,P(x-1,\dots,x+n-1,t).
\end{align}
It is seen that the rate of particles all hopping to right is
simply the rate of one particle hopping to right, to the power of
the the number of particles. This shows that there is no pushing.
This is the simple exclusion process.

For the second case, one obtains
\begin{align}\label{f8}
P(x,x+1,\dots,x+n,t+1)=&(1-\alpha)^{n+1}\,P(x,\dots,x+n,t)\nonumber\\
&+\alpha\sum_{m=0}^n(1-\alpha)^{n-m}\nonumber\\
&\times P(\dots,x+m-1,x+m+1,\dots,x+n,t).
\end{align}
This shows that there is a pushing process, the probability of
which does not depend on the length of the block:
\begin{equation}\label{f9}
A\cdots A\emptyset\to\emptyset A\cdots A,\qquad\hbox{with the
probability }\alpha.
\end{equation}
This is the drop-push model.

\section{The conditional probability}
The $n$-particle analogue of (\ref{f2}), can be written as
\begin{align}\label{f10}
P(\mathbf{x},t+1)&=(UP)(\mathbf{x},t),\nonumber\\
&=[(1-\alpha+\alpha\,T_1)\cdots(1-\alpha+\alpha\,T_n)P](\mathbf{x},t),
\end{align}
where
\begin{equation}\label{f11}
(T_jP)(x_1,\dots,x_n,t):=P(x_1,\dots,x_j-1,\dots,x_n,t).
\end{equation}
For the evolution equation (\ref{f10}), the Bethe-ansatz solution
(the eigenvector of $U$) corresponding to the eigenvalue $u$ is
\begin{equation}\label{f12}
u\,\Psi(\mathbf{x})=[(1-\alpha+\alpha\,T_1)\cdots(1-\alpha+\alpha\,T_n)\Psi](\mathbf{x}),
\end{equation}
subject to the condition
\begin{align}\label{f13}
\Psi(\dots,x_j=x,x_{j+1}=x,\dots)=&\lambda\,\Psi(\dots,x_j=x,x_{j+1}=x+1,\dots)\nonumber\\
&+\mu\,\Psi(\dots,x_j=x-1,x_{j+1}=x,\dots).
\end{align}
Using the Bethe-ansatz
\begin{equation}\label{f14}
\Psi_{\mathbf{k}}(\mathbf{x})=\sum_{\sigma}A_\sigma
e^{i\,\mathbf{x}\cdot\sigma(\mathbf{k})},
\end{equation}
where $\sigma$ runs over $n$-permutations and
\begin{equation}\label{f15}
A_1=1,
\end{equation}
one arrives at
\begin{equation}\label{f16}
u=\prod_{j=1}^n(1-\alpha+\alpha\,e^{-i\,k_j}),
\end{equation}
and
\begin{equation}\label{f17}
A_{\sigma\,\sigma_j}=S(k_{\sigma(j)},k_{\sigma(j+1)})\,A_\sigma,
\end{equation}
where $\sigma_j$ changes $j$ to $j+1$ and $j+1$ to $j$, and leaves
the other numbers between 1 and $n$ intact, and
\begin{equation}\label{f18}
S(k_1,k_2)=-\frac{1-\lambda\,e^{i\,k_2}-\mu\,e^{-i\,k_1}}{1-\lambda\,e^{i\,k_1}-\mu\,e^{-i\,k_2}}.
\end{equation}
This derivation is essentially the same as that used in
\cite{AKK1,AKK2,RK}.

Using these, the conditional probability (of finding the particles
at $\mathbf{x}$ at the $t$, when they have been at $\mathbf{y}$ at
the 0) is obtained as
\begin{equation}\label{f19}
P(\mathbf{x},t;\mathbf{y},0)=\int\frac{\mathrm{d}^nk}{(2\pi)^n}\;\Psi_{\mathbf{k}}(\mathbf{x})\,
e^{-i\,\mathbf{k}\cdot\mathbf{y}}\,u^t,
\end{equation}
where the integration runs from 0 to $2\pi$, for each of the
$k_j$'s. Also, to treat the singularity arising from $S$ in
$\Psi$, one is supposed to multiply $\lambda$ and $\mu$ in the
denominator by $e^{-\epsilon}$. It is seen that the right-hand
side is equal $\delta_{\mathbf{x},\mathbf{y}}$ (for $\mathbf{x}$
and $\mathbf{y}$ in the physical region). So, the right-hand side
satisfies the appropriate initial condition and evolution equation
for the conditional probability, and hence is the (unique)
solution to the conditional probability.

\section{The drift rates}
In the two-particle sector, the one-particle probabilities are
defined as
\begin{align}\label{f20}
P_1(x,t):=&\sum_{x_2>x}P(x,x_2,t)\nonumber\\
P_2(x,t):=&\sum_{x_1<x}P(x_1,x,t).
\end{align}
From these,
\begin{align}\label{f21}
P_1(x,t+1)=&(1-\alpha)\,P_1(x,t)+\alpha\,P_1(x-1,t)\nonumber\\
&+\lambda\,\alpha\,(1-\alpha)\,[P(x,x+1,t)-P(x-1,x,t)],\nonumber\\
P_2(x,t+1)=&(1-\alpha)\,P_2(x,t)+\alpha\,P_2(x-1,t)\nonumber\\
&+ \mu\,\alpha\,(1-\alpha)\,[P(x-2,x-1,t)-P(x-1,x,t)].
\end{align}
Defining
\begin{align}\label{f22}
\langle X_1\rangle(t):=&\sum_x x\,P_1(x,t),\nonumber\\
\langle X_2\rangle(t):=&\sum_x x\,P_2(x,t),
\end{align}
(the expectation value of the position of the first and second
particles) one has
\begin{align}\label{f23}
\langle X_1\rangle(t+1)=&\langle X_1\rangle(t)+\alpha-\lambda\,\alpha\,(1-\alpha)\,P_r(1,t),\nonumber\\
\langle X_2\rangle(t+1)=&\langle
X_2\rangle(t)+\alpha+\mu\,\alpha\,(1-\alpha)\,P_r(1,t),
\end{align}
where
\begin{equation}\label{f24}
P_r(x,t):=\sum_y P(y,y+x,t).
\end{equation}
Writing (\ref{f19}) for the two-particle sector,
\begin{align}\label{f25}
P(x_1,x_2,t;y_1,y_2,0)=\int\frac{\mathrm{d}^2k}{(2\pi)^2}\;
\Big[&e^{i\,(k_1\,x_1+k_2\,x_2)}\nonumber\\
&-
\frac{1-\lambda\,e^{i\,k_2}-\mu\,e^{-i\,k_1}}{1-\lambda\,e^{i\,k_1}-\mu\,e^{-i\,k_2}}\,
e^{i\,(k_2\,x_1+k_1\,x_2)}\Big]\,e^{-i\,(k_1\,y_1+k_2\,y_2)}\nonumber\\
&\times(1-\alpha+\alpha\,e^{-i\,k_1})^t\,(1-\alpha+\alpha\,e^{-i\,k_2})^t,
\end{align}
one arrives at
\begin{align}\label{f26}
P_{r}(x,t)=&\int\frac{\mathrm{d}k}{2\pi}\,
[e^{i\,k\,x}+e^{-i\,k\,(x-1)}]\,e^{-i\,k\,(y_2-y_1)}\nonumber\\
&\times(1-\alpha+\alpha\,e^{-i\,k})^t\,(1-\alpha+\alpha\,e^{i\,k})^t,
\end{align}
where, using (\ref{f24}), the summation over $y$ is done, which
leads to a delta function $\delta(k_1+k_2)$, using which one of
the integrations is carried out.

A steepest descent calculation shows that if $t$ is large and $x$
is not large,
\begin{equation}\label{f27}
P_r(x,t)\sim\frac{1}{\sqrt{\pi\,\alpha\,(1-\alpha)\,t}}.
\end{equation}
So, for large $t$,
\begin{align}\label{f28}
\langle X_1\rangle(t)=&\langle X_1\rangle(0)+\alpha\,t-\lambda\,\left[2\sqrt{\frac{\alpha\,(1-\alpha)\,t}{\pi}}+C+o(1)\right],\nonumber\\
\langle X_2\rangle(t)=&\langle
X_2\rangle(0)+\alpha\,t+\mu\,\left[2\sqrt{\frac{\alpha\,(1-\alpha)\,t}{\pi}}+C+o(1)\right].
\end{align}
One also has
\begin{align}\label{f29}
(\langle X_2\rangle-\langle X_1\rangle)(t+1)=&(\langle X_2\rangle-\langle X_1\rangle)(t)+
\alpha\,(1-\alpha)\,P_r(1,t),\nonumber\\
\langle X\rangle(t+1)=&\langle
X\rangle(t)+\alpha+\frac{\mu-\lambda}{2}
\,\alpha\,(1-\alpha)\,P_r(1,t),\nonumber\\
(\mu\,\langle X_1\rangle+\lambda\,\langle
X_2\rangle)(t+1)=&(\mu\,\langle X_1\rangle+\lambda\,\langle
X_2\rangle)(t)+\alpha,
\end{align}
where
\begin{equation}\label{f30}
\langle X\rangle:=\frac{1}{2}\,(\langle X_1\rangle+\langle
X_2\rangle),
\end{equation}
is the expectation value of the position of the particles. So for
all times
\begin{equation}\label{f31}
(\mu\,\langle X_1\rangle+\lambda\,\langle
X_2\rangle)(t)=(\mu\,\langle X_1\rangle+\lambda\,\langle
X_2\rangle)(0)+\alpha\,t,
\end{equation}
and for large times,
\begin{align}\label{f32}
(\langle X_2\rangle-\langle X_1\rangle)(t)=&(\langle X_2\rangle-\langle X_1\rangle)(0)+2\sqrt{\frac{\alpha\,(1-\alpha)\,t}{\pi}}+C+o(1),\nonumber\\
\langle X\rangle(t)=&\langle X\rangle(0)+\alpha\,t\nonumber\\
&+(\mu-\lambda)\,\left[\sqrt{\frac{\alpha\,(1-\alpha)\,t}{\pi}}+C+o(1)\right],
\end{align}
where $C$ is a constant. So the drift rates are large $t$ are
\begin{align}\label{f33}
V_1:=\frac{\mathrm{d}\langle X_1\rangle}{\mathrm{d}t}=&\alpha-\lambda\,\sqrt{\frac{\alpha\,(1-\alpha)}{\pi\,t}},\nonumber\\
V_2:=\frac{\mathrm{d}\langle
X_2\rangle}{\mathrm{d}t}=&\alpha+\mu\,\sqrt{\frac{\alpha\,(1-\alpha)}{\pi\,t}}.
\end{align}
$\langle X_1\rangle$ and $\langle X_2\rangle$ are the expectation
values of the positions of the first and second particles,
respectively, and $V_1$ and $V_2$ are their corresponding
velocities. As $t$ is discrete, these velocities are defined only
when the $X_i$'s are smooth functions of $t$, which happens at
large times.

The above equations show that the drift velocities of both
particles approach $\alpha$ for large times. The reason is that at
large times the particles are far from each other and effectively
do not interact with each other. But the next leading terms in
velocities are negative for the first particle and positive for
the second particle, which is expected from the hindering effect
of the second particle on the first, and the pushing effect of the
first particle on the second. One can see that the results
obtained in \cite{RK} are recovered, provided one replaces
$\alpha\,(1-\alpha)\,t$ with $t$.

\section{The diffusion rate}
Starting from (\ref{f21}), and defining
\begin{align}\label{f34}
\langle X^2_1\rangle(t):=&\sum_x x^2\,P_1(x,t),\nonumber\\
\langle X^2_2\rangle(t):=&\sum_x x^2\,P_2(x,t),
\end{align}
one has
\begin{align}\label{f35}
\langle X^2_1\rangle(t+1)=&\langle X^2_1\rangle(t)+2\alpha\,
\langle X_1\rangle(t)+\alpha\nonumber\\
&-\lambda\,\alpha\,(1-\alpha)\sum_x(2x+1)\,P(x,x+1,t),\nonumber\\
\langle X^2_2\rangle(t+1)=&\langle X^2_2\rangle(t)+2\alpha\,
\langle X_2\rangle(t)+\alpha\nonumber\\
&+\mu\,\alpha\,(1-\alpha)\sum_x(2x+3)\,P(x,x+1,t).
\end{align}
Defining
\begin{align}\label{f36}
\langle X^2\rangle:=&\frac{1}{2}\,(\langle X^2_1\rangle+\langle
X^2_2\rangle),\nonumber\\
\Delta^2:=&\langle X^2\rangle-(\langle X\rangle)^2,
\end{align}
($\Delta^2$ is the variance of the position of the particles) one
arrives at
\begin{align}\label{f37}
\Delta^2(t+1)=&\Delta^2(t)+\alpha\,(1-\alpha)\nonumber\\
&+\alpha\,(1-\alpha)\,(\mu-\lambda)\left[\frac{1}{2}\,\sum_x(2x+1)\,P(x,x+1,t)
-P_r(1,t)\,\langle X\rangle(t)\right]\nonumber\\
&+\alpha\,(1-\alpha)\,[\mu-\alpha\,(\mu-\lambda)]\,P_r(1,t)\nonumber\\
&-\frac{\alpha^2\,(1-\alpha)^2\,(\mu-\lambda)^2}{4}\,P_r^2(1,t).
\end{align}
From (\ref{f27}), it is seen that the last two terms on the
right-hand side vanish as $t\to\infty$. One also has
\begin{align}\label{f38}
\sum_x(2x+1)\,P(x,x+1,t)=&\sum_x\int\frac{\mathrm{d}^2k}{4\pi^2}\,(2x+1)\,
e^{i(k_1+k_2)\, x}\nonumber\\
&\times\left[e^{i\,k_2}-\frac{1-\lambda\,e^{i\,k_2}-\mu\,e^{-i\,k_1}}
{1-\lambda\,e^{i\,k_1}-\mu\,e^{-i\,k_2}}\,e^{i\,k_1}\right]\,
e^{-i\,(k_1\,y_1+k_2\,y_2)}\nonumber\\
&\times
(1-\alpha+\alpha\,e^{-i\,k_1})^t\,(1-\alpha+\alpha\,e^{-i\,k_2})^t,\nonumber\\
=&\int\frac{\mathrm{d}^2k}{4\pi^2}\,[2\pi\,\delta(k_1+k_2)-4\pi\,i\,\delta'(k_1+k_2)]
\nonumber\\
&\times\left[e^{i\,k_2}-\frac{1-\lambda\,e^{i\,k_2}-\mu\,e^{-i\,k_1}}
{1-\lambda\,e^{i\,k_1}-\mu\,e^{-i\,k_2}}\,e^{i\,k_1}\right]\,e^{-i\,(k_1\,y_1+k_2\,y_2)}
\nonumber\\
&\times
(1-\alpha+\alpha\,e^{-i\,k_1})^t\,(1-\alpha+\alpha\,e^{-i\,k_2})^t.
\end{align}
Here $\delta'$ is the derivative of $\delta$ with respect to its
argument. We are seeking those terms on the right-hand side of
(\ref{f37}), which don't vanish as $t\to\infty$. On the right-hand
side of (\ref{f38}), it is seen that the terms proportional to
$\delta(k_1+k_2)$ in the integrand, give rise to terms
proportional to $t^{-1/2}$ (for large $t$). (In fact these terms
are proportional to $P_r(1,t)$.) Denoting the terms coming from
that part of the integrand which is proportional to
$\delta'(k_1+k_2)$ by $I$, one has
\begin{align}\label{f39}
I=&\int\frac{i\,\mathrm{d}^2k}{\pi}\,\delta(k_1+k_2)\,
\frac{\partial}{\partial k_2}\,\Big\{
\Big[e^{i\,k_2}-\frac{1-\lambda\,e^{i\,k_2}-\mu\,e^{-i\,k_1}}
{1-\lambda\,e^{i\,k_1}-\mu\,e^{-i\,k_2}}\,e^{i\,k_1}\Big]\nonumber\\
&\times e^{-i\,(k_1\,y_1+k_2\,y_2)}\,
(1-\alpha+\alpha\,e^{-i\,k_1})^t\,(1-\alpha+\alpha\,e^{-i\,k_2})^t\Big\}.
\end{align}
Differentiation of the fraction, $S(k_1,k_2)$, needs some care.
The derivative of this fraction is singular at $k_1=k_2=0$. To
remove this singularity, one has to replace $k_1$ with
$k_1+i\,\epsilon$ in the denominator. This prescription guaranties
that in (\ref{f25}), the integral of the second term on the
right-hand side tends to zero as $x_2$ tends to infinity. We are
interested in the behavior of $I$ for large times, and this is
determined by the behavior of the integrand multiplier of
$\delta(k_1+k_2)$ for $k_1=-k_2$, where $k_2$ is small. One has
\begin{align}\label{f40}
\frac{\partial}{\partial k_2}
\left(\frac{1-\lambda\,e^{i\,k_2}-\mu\,e^{i\,k_1}}
{1-\lambda\,e^{i\,k_1}-\mu\,e^{i\,k_2}}\right)
\Big|_{k_1=-k_2=-k\sim 0}=&-i\,
\frac{\lambda\,e^{i\,k}-\mu}{1-e^{i\,k}}\Big|_{k\sim 0}, \nonumber\\
=&\frac{\mu-\lambda}{k-i\,\epsilon}\Big|_{k\sim 0}, \nonumber\\
=&(\mu-\lambda)\,\mathrm{pf}\left(\frac{1}{k}\right)+i\,\pi\,(\mu-\lambda)\,\delta(k).
\end{align}
Here $\mathrm{pf}$ means a pseudo-function (the Cauchy principal
value in integration). Putting this in (\ref{f39}), and keeping
only terms which don't vanish as $t\to\infty$, one arrives at
\begin{align}\label{f41}
I=&\int\frac{i\,\mathrm{d}k}{\pi}[-i\,\pi\,(\mu-\lambda)\,
\delta(k)+2(-i\,\alpha\,t)\,(1-\alpha+\alpha\,e^{i\,k})^t\,
(1-\alpha+\alpha\,e^{-i\,k})^t]+\cdots,\nonumber\\
=&(\mu-\lambda)+\frac{2\alpha\,t}{\sqrt{\pi\,\alpha\,(1-\alpha)\,t}}+\cdots.
\end{align}
So,
\begin{equation}\label{f42}
\sum_x(2x+1)\,P(x,x+1,t)=(\mu-\lambda)+
\frac{2\alpha\,t}{\sqrt{\pi\,\alpha\,(1-\alpha)\,t}}+\cdots.
\end{equation}
Using (\ref{f27}) and (\ref{f32}), one arrives at
\begin{equation}\label{f43}
P_r(1,t)\,\langle X\rangle(t)=\frac{\mu-\lambda}{\pi}
+\frac{\alpha\,t}{\sqrt{\pi\,\alpha\,(1-\alpha)\,t}}+\cdots.
\end{equation}
Putting (\ref{f42}) and (\ref{f43}) in (\ref{f37}), one arrives at
\begin{equation}\label{f44}
\Delta^2(t+1)=\Delta^2(t)+\alpha\,(1-\alpha)\,\left[1+(\mu-\lambda)^2\,
\left(\frac{1}{2}-\frac{1}{\pi}\right)\right]+\cdots,
\end{equation}
from which,
\begin{equation}\label{f45}
\lim_{t\to\infty}\frac{\mathrm{d}\Delta^2}{\mathrm{d}t}=\alpha\,(1-\alpha)\,
\left[1+(\mu-\lambda)^2\,
\left(\frac{1}{2}-\frac{1}{\pi}\right)\right].
\end{equation}
This diffusion rate, is again defined only at large times, when
one can treat $\Delta^2$ as a function of continuous time. It is
seen that it is in agreement with that obtained in \cite{RK},
provided one replaces $\alpha\,(1-\alpha)\,t$ with $t$.
\section{concluding remarks}
The main result of the paper was to introduce a discrete-time
discrete-space model, solvable through the Bethe-ansatz method.
The model contains a free parameter (say $\lambda$) that for
certain values reproduces the simple exclusion model and the
drop-push model. The conditional probability of finding particles
at different sites was obtained as a function of time, from which
in principle one can derive any correlation function. The
conditional probabilities were calculated for the general
multi-particle sector. The drift- and diffusion-rates, however,
were explicitly calculated only for the two-particle sector, and
it was shown that the results agreed with those of the
continuous-time system at large-times. There remains the question
of performing similar calculations for the multi-particle sector.
For large times, one can put forward the following arguments. For
large times, only the behavior of the integrand in (\ref{f19})
around $\mathbf{k}=\mathbf{0}$ is important, and it is seen that
for $\lambda=\mu$, there is no pole in the scattering matrix $S$.
In fact, $S$ becomes one for $\lambda=\mu$ and
$\mathbf{k}=\mathbf{0}$, which shows that for $\lambda=\mu$ and
for large times, the conditional probability takes the form
\begin{equation}\label{f46}
P(\mathbf{x},t;\mathbf{y},0)=\sum_\sigma
P_0[\sigma(\mathbf{x}),t;\mathbf{y},0],
\end{equation}
where $P_0$ is the conditional probability of a system consisting
of free particles hopping to the right, which corresponds to the
one obtained with $S=0$ in (\ref{f19}). This shows that for
$\lambda=\mu$, at large times the system behaves collectively as a
collection of free particles. Hence the drift- and diffusion-rates
should be $\alpha$ and $\alpha(1-\alpha)$ respectively, which
agrees with the particular case of the two-particles.

For the general case $\lambda\ne\mu$, at large times the particles
will generally be far from each other. The interaction terms
coming from the scattering matrix, are in the from of products of
two-particle scattering matrices. So it is plausible that for
large times and for calculating up to 2-point functions, one
neglects more-than two-particle interactions and interactions
between non-adjacent particles. Then, from the $n!$ terms in the
Bethe-ansatz solution for the $n$-particle sector, there remains
only $n$ terms. The drift rate at large times is expected to
remain $\alpha$ again. For the diffusion rate, one can argue that
it should be $\alpha(1-\alpha)$ (the free-particle value) plus a
function of $(\mu-\lambda)$ which vanishes at $\mu=\lambda$. The
additional term comes from the interaction of the neighboring
particles. So it is expected to be proportional to
$(\lambda-\mu)^2$. This shows that the drift- and diffusion-rates
obtained for the two-particle sector at large times, serve as
qualitative results for the multi-particle sector as well.

\end{document}